%% file: main.tex
\documentclass[11pt, french, apa]{article}

\usepackage{lipsum}

\usepackage{graphicx} % insérer des images
\usepackage{svg} % insérer des graphiques SVG
\usepackage{xcolor}
\definecolor{accent}{RGB}{69, 165, 157}
\definecolor{darkaccent}{RGB}{66, 74, 84}

\usepackage[french]{babel}

\usepackage[hidelinks]{hyperref}
\hypersetup{pdfauthor={Plaisant Oscar, Lemoine Max}, pdftitle={Pourquoi existe-t-il de nombreux paradigmes de programmation ?}}

\usepackage{enumitem}
\setlist[itemize]{label=\Large\raisebox{-0.2ex}\textbullet}

\usepackage{ragged2e} % justification

\usepackage[top=2.5cm,
            bottom=2.5cm,
            left=2.5cm,
            right=2.5cm]{geometry}
\usepackage[skip=6pt]{parskip}

\usepackage{titlesec}

\titleformat{\section}{%
    \color{accent}\normalfont\LARGE}{\thesection}{1em}{}
\titlespacing{\section}{0pt}{12pt}{12pt}

\titleformat{\subsection}{\color{darkaccent}\normalfont\Large}{\thesubsection}{1em}{}
\titlespacing{\subsection}{0pt}{12pt}{12pt}

\titleformat{\subsubsection}{\color{darkaccent}\normalfont\large}{\thesubsubsection}{1em}{}

\titleformat*{\paragraph}{\bfseries\color{darkaccent}}

\usepackage{titletoc}
\dottedcontents{section}[0em]{}{2ex}{6pt}
\dottedcontents{subsection}[3em]{}{4ex}{6pt}
\dottedcontents{subsubsection}[6em]{}{6ex}{6pt}

\usepackage[format=plain,
            font={color=darkaccent, it}]{caption}
\addto\captionsfrench{}
\addto\captionsfrench{}

\newtheorem{definition}{Definition}%[section]

\usepackage[natbibapa]{apacite} % citations au format APA

\begin{document}
% changer le style du titre de la table des matières
\renewcommand{\contentsname}{\Large\textcolor{black}{\bfseries Table des matières}}
% justifier le texte
\justifying

\input{src/titlepage.tex}

\tableofcontents

\newpage

\input{tex_files/0-introduction}

\newpage

\input{tex_files/1-definitions}

\input{tex_files/2-apprentissage}

\input{tex_files/3-resolution-problemes}

\input{tex_files/4-conclusion}

\newpage

\input{tex_files/9-annexes}

\newpage

\bibliographystyle{apacite}
\bibliography{main.bib}

\end{document}

%% file: src/titlepage.tex
\begin{titlepage}
    \begin{center}
        %\includesvg{img/logo_universite.svg}
        \includegraphics{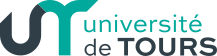}

        \vspace{3cm}
        % Titre
        % {\Huge\bfseries Influence des paradigmes de programmation sur l'efficacité et l'efficience pour la résolution de problèmes par des humains}
        %{\Huge\bfseries test}
        {\Huge\bfseries Pourquoi existe-t-il de nombreux paradigmes de programmation ?}

        \vspace{6.5cm minus 4.5cm}
        % Auteurs
        {\Large
        \textbf{Plaisant Oscar}
        
        \textbf{Lemoine Max}
        }

        \vspace{3cm minus 1cm}
        % encadrant
        Enseignante encadrante : Markhoff Béatrice
        
        \vspace{3cm minus 1cm}

        \textbf{Licence Informatique L3}

        \textbf{Promotion 2021-2024}

    \end{center}
\end{titlepage}

%% file: tex_files/0-introduction.tex
\section{Introduction}\label{introduction}

Il existe de nombreux paradigmes de programmation.
Puisque tous les langages Turing-complets sont formellement équivalents (ils ont la même capacité à exprimer l'ensemble des problèmes calculables), l'existence de tant de paradigmes différents peut sembler étonnante, voire inutile.
Nous essayerons de comprendre pourquoi il existe tant de paradigmes différents.
Nous présenterons d'abord une définition de ce qu'est un paradigme de programmation, puis nous exposerons en quoi différents paradigmes sont plus adaptés pour différentes raisons : pour l'apprentissage, pour la résolution ou l'expression de certains types de problèmes et pour les apports que fait chaque paradigme en général.

%% file: tex_files/1-definitions.tex
\section{Définitions et concepts
importants}\label{definition-et-concepts-importants}

\subsection{Définition de concepts
importants}\label{definition-de-concepts-importants}

\subsubsection{État}\label{etat}

L'état est la capacité, pour un programme, à retenir de l'information.
``State is the ability to remember information, or more precisely, to store a sequence of values in time.'' (L'état est la capacité à retenir de l'information, ou, plus précisément, à stocker une séquence de valeurs dans le temps) \citep[p. 14]{royProgrammingParadigmsDummies}.

Il est important de noter que la présence de variables n'implique pas toujours la présence d'état.
En effet, une fonction peut avoir un paramètre sans avoir nécessairement d'état.
Les variables des fonctions, si elles constituent une référence à une valeur, n'impliquent pas la présence d'état \footnote{%
    La logique combinatoire fournit la preuve qu'il est possible d'établir un système de calcul sans aucune variable ni référence, ce qui rend l'absence d'état évidente \citep{LogiqueCombinatoire2023}.}.

\subsubsection{Fonction et procédure}\label{fonction-et-procedure}

Les termes ``fonction'' et ``procédure'' sont, selon les contextes et les auteurs, utilisés pour désigner différentes choses.

Par exemple, le livre ``Structure and interpretation of computer programs'' \citep{abelsonStructureInterpretationComputer1996} fournit la distinction suivante : les fonctions sont l'aspect mathématique du concept et sont définies en déclarant ``ce qu'elle est'', en donnant des propriétés de cette fonction (une définition déclarative, voir \ref{programmation-declarative}).
Les procédures sont l'aspect programmatique du concept, et sont définies en déclarant ``comment elle doit faire'', en décrivant comment elle doit s'exécuter \footnote{%
    ``The contrast between function and procedure is a reﬂection of the general distinction between describing properties of things and describing how to do things, or, as it is sometimes referred to, the distinction between declarative knowledge and imperative knowledge.
In mathematics we are usually concerned with declarative (what is) descriptions, whereas in computer science we are usually concerned with imperative (how to) descriptions.'' \citep[p. 28]{abelsonStructureInterpretationComputer1996}}.

Cependant, certains auteurs utilisent le terme \emph{procedure} même pour parler du concept théorique et mathématique (qui se rapproche du lambda-calcul) \footnote{%
    Par exemple, Fellesein dans cette expression : ``(let x be v in e) is expressible as (apply (procedure x e) v)'' \citep[p. 135]{felleisenExpressivePowerProgramming1990}}.

Nous utiliserons donc principalement le terme fonction, avec les définitions suivantes :

\begin{definition}
    Une fonction est l'encapsulation d'un ensemble d'instructions.
    Une fonction peut posséder des paramètres qui peuvent influencer son exécution.
    Ces paramètres sont des valeurs ou des références qui sont données (``passées en argument'') lors de l'appel de cette fonction.
    Une fonction peut retourner une valeur.
    Une fonction peut être appelée plusieurs fois et à plusieurs endroits, ce qui la rend réutilisable dans plusieurs contextes.
\end{definition}

\begin{definition}
    Une fonction pure est une fonction déterministe, c'est-à-dire que les mêmes arguments donnent toujours la même valeur de retour.
    Une fonction pure est une fonction sans effet de bord, c'est-à-dire qu'elle ne modifie pas l'état du système (en dehors de son champ local propre, dont la durée de vie est limitée).
\end{definition}

\subsubsection{Effet de bord}\label{effet-de-bord}

\begin{definition}
    Un effet de bord est la modification d'état par une fonction.
\end{definition}

Du point de vue du développeur, cela arrive lorsqu'une fonction effectue des modifications en dehors de son propre corps.
Cela peut être fait en utilisant des références ou des constructions particulières du langage.

Voici des exemples de mécanismes qui permettent des effets de bord \footnote{%
    ``Par exemple, les fonctions qui modifient une variable locale statique, une variable non locale ou un argument mutable passé par référence, les fonctions qui effectuent des opérations d'entrées-sorties ou les fonctions appelant d'autres fonctions à effet de bord'' \citep{EffetBordInformatique2023}} :

\begin{itemize}
    \item  Les variables globales : en permettant à toutes les fonctions d'accéder à un même champ de mémoire, les variables globales leur donnent la possibilité de modifier l'état du système 
    \item Les variables statiques locales : en permettant aux objets d'une même classe de partager une variable, on leur permet de générer des effets de bord contrôlés 
    \item Les arguments mutables (les arguments passés par référence) : ce mécanisme permet de répercuter la modification d'un paramètre en modification de l'argument, ce qui permet à la fonction de modifier une valeur qui lui a été passée en argument. Cela constitue un effet de bord
\end{itemize}

\subsubsection{Fermeture}\label{fermeture}

Une fermeture (de l'anglais \emph{closure}) est une fonction à laquelle on attache l'environnement local de l'endroit où elle a été définie \footnote{%
    ``From an implementation viewpoint, a closure combines a procedure with its external references (the references it uses at its deﬁnition).'' \citep[p. 24]{royProgrammingParadigmsDummies}}.
Cela permet à cette fonction d'accéder aux valeurs et aux références de cet environnement local, même si elle est appelée depuis un autre contexte (dans un autre champ local) \footnote{%
    ``Dans un langage de programmation, une fermeture ou clôture (en anglais~: closure) est une fonction accompagnée de son environnement lexical. L'environnement lexical d'une fonction est l'ensemble des variables non locales qu'elle a capturées, soit par valeur (c'est-à-dire par copie des valeurs des variables), soit par référence (c'est-à-dire par copie des adresses mémoires des variables){[}1{]}. Une fermeture est donc créée, entre autres, lorsqu'une fonction est définie dans le corps d'une autre fonction et utilise des paramètres ou des variables locales de cette dernière.'' \citep{FermetureInformatique2024}}.
Du point de vue du développeur, cela signifie que l'exécution d'une fermeture soit le même que si les instructions étaient exécutées ``à l'endroit'' (dans le contexte où elle a été définie) la fermeture a été créée \footnote{%
    From the programmer's viewpoint, a closure is a ``packet of work'': a program can transform any instructions into a closure at one point in the program, pass it to another point, and decide to execute it at that point. The result of its execution is the same as if the instructions were executed at the point the closure was created. \citep[p. 24]{royProgrammingParadigmsDummies}}.

Un avantage considérable des fermetures est qu'elles permettent d'accéder aux valeurs et références d'un environnement local qui n'existe plus (car il a été libéré de la pile d'exécution) \footnote{%
    ``Une fermeture peut être passée en argument d'une fonction dans l'environnement où elle a été créée (passée vers le bas) ou renvoyée comme valeur de retour (passée vers le haut). Dans ce cas, le problème posé alors par la fermeture est qu'elle fait référence à des données qui auraient typiquement été allouées sur la pile d'exécution et libérées à la sortie de l'environnement. Hors optimisations par le compilateur, le problème est généralement résolu par une allocation sur le tas de l'environnement.'' \citep{FermetureInformatique2024}}.

\subsection{Qu'est-ce qu'un paradigme ?}\label{quest-ce-quun-paradigme}

Un paradigme de programmation est une façon d'approcher la programmation et de formuler les problèmes et leur formalisation dans un langage de programmation \footnote{%
    ``Le paradigme de programmation est la façon (parmi d'autres) d'approcher la programmation informatique et de formuler les solutions aux problèmes et leur formalisation dans un langage de programmation approprié. Ce n'est pas de la méthodologie contenant une méthode~; cette dernière organise le traitement des problèmes reconnus dans l'écosystème concerné pour aboutir à la solution conceptuelle et programme exécutable.'' \citep{ParadigmeProgrammation2023}}.
En particulier, un paradigme fournit et détermine comment un développeur doit voir un programme. La notion de paradigme est notamment à dissocier de celles de \emph{méthode} ou bien de \emph{design patterns}, qui décrivent comment traiter des problèmes spécifiques et reconnus, et comment aboutir à une solution conceptuelle \citep{ParadigmeProgrammation2023}. Un paradigme est un concept plus ``haut niveau'', c'est-à-dire plus abstrait : chaque paradigme supporte un ensemble particulier de concepts (cohérents entre eux), qui peuvent être hérités d'une théorie mathématique, de principes fondamentaux, ou bien d'une vision sur ce que doit être la programmation \footnote{%
    ``A programming paradigm is an approach to programming a computer based on a mathematical theory or a coherent set of principles.'' \citep{royProgrammingParadigmsDummies}}.

Un paradigme de programmation est souvent principalement décrit par les concepts qu'il implémente ou non \footnote{%
    ``Each paradigm supports a set of concepts that makes it the best for a certain kind of problem. For example, object-oriented programming is best for problems with a large number of related data abstractions organized in a hierarchy. Logic programming is best for transforming or navigating complex symbolic structures according to logical rules. Discrete synchronous programming is best for reactive problems, i.e., problems that consist of reactions to sequences of external events.'' \citep[p. 10]{royProgrammingParadigmsDummies}}.
C'est notamment cette approche qu'emploie la taxonomie des paradigmes de programmation définie par Peter Van Roy \citep{TaxonomiePrincipauxParadigmes}.

Cependant, J.Huges, dans son article ``Why Functional Programming Matters'' \citep[p. 1]{hughesWhyFunctionalProgramming1989}, fustige le fait que certains paradigmes (particulièrement la programmation fonctionnelle (voir la section \ref{programmation-fonctionnelle}) et la programmation structurée (voir la section \ref{programmation-structuree})) sont trop fréquemment définis en termes des fonctionnalités qu'ils n'implémentent pas, ou des contraintes qu'ils posent \footnote{%
    ``It says a lot about what functional programming is~\emph{not}~{[}\ldots{]} but not much about what it is.'' \citep[p. 1]{hughesWhyFunctionalProgramming1989}}.
Cela est problématique, car l'absence d'une fonctionnalité ne peut pas expliquer pourquoi un paradigme serait plus intéressant dans certains cas \footnote{%
    ``It is a logical impossibility to make a language more powerful by ommitting features, no matter how bad they may be'' \citep[p. 1]{hughesWhyFunctionalProgramming1989}}.
L'auteur insiste donc sur le fait que les paradigmes devraient être définis en fonction des avantages structurels qu'ils apportent pour la résolution de certains types de problèmes.

C'est pour cette raison que, dans notre définition des différents
paradigmes, nous chercherons à expliquer à la fois les principes
fondamentaux de chaque paradigme, mais également à montrer brièvement
pourquoi ceux-ci sont intéressants, c'est-à-dire répondre à la question
``en quoi ce paradigme apporte-t-il quelque chose ?''.

Cela constituera une première partie de réponse à notre problématique.

\subsection{Description de quelques paradigmes}\label{description-de-quelques-paradigmes}

\subsubsection{Programmation impérative}\label{programmation-imperative}

La programmation impérative est un paradigme de programmation dans lequel le contrôle de flot d'exécution est explicite.
Cela signifie que le programme spécifie pas à pas la marche à suivre pour obtenir le bon résultat.
En programmation impérative, chaque étape modifie l'état global du système \footnote{% 
    ``Control flow in imperative programming is explicit: commands show how the computation takes place, step by step. Each step affects the global state of the computation.'' \citep{toalProgrammingParadigms}}.

\subsubsection{Programmation déclarative}\label{programmation-declarative}

La programmation déclarative est un paradigme de programmation dans lequel la définition des programmes se fait en déclarant la forme du résultat plutôt que la manière l'obtenir (comme le fait la programmation impérative, voir section \ref{programmation-imperative}).
Le développeur ne s'occupe donc pas de l'exécution, mais plutôt de comment spécifier le résultat.
Le contrôle du flot d'exécution est implicite.
En cela, la programmation déclarative s'oppose à la programmation impérative \footnote{%
    ``Control flow in declarative programming is implicit: the programmer states only what the result should look like, not how to obtain it.'' \citep{toalProgrammingParadigms}}.

Un avantage évident de la programmation déclarative est qu'elle libère le programmeur de certaines charges mentales, notamment celle de spécifier l'ordre d'exécution, le flot de contrôle de l'exécution.
Cela rend le paradigme déclaratif plus pratique pour le développeur (plus haut niveau), puisqu'il ne requiert pas de s'occuper de certains détails qui n'influent pas sur le sens du programme \footnote{%
    ``A programming language is low level when its programs require attention to the irrelevant.'' \citep{perlisSpecialFeatureEpigrams1982}}.

\subsubsection{Programmation fonctionnelle}\label{programmation-fonctionnelle}

La programmation fonctionnelle est un paradigme de programmation dans lequel les programmes sont exprimés comme des arbres d'expressions.

Le contrôle du flot d'exécution est donc fait en combinant des fonctions \footnote{%
    ``In functional programming, control flow is expressed by combining function calls, rather than by assigning values to variables'' \citep{toalProgrammingParadigms}} :
la fonction principale prend en argument l'entrée du problème, puis elle est définie à partir d'autres fonctions, qui sont elles-mêmes définies à partir de fonctions, et ce, jusqu'à ce que les fonctions à la base de la définition (les feuilles de l'arbre d'expressions) soient des primitives du langage, ou bien des constantes \footnote{%
    ``Functional programming is so called because a program consists entirely of functions. The main program itself is written as a function which recieves the program's input as its argument and delivers the program's output as its result. Typically the main function is defined in terms of other functions, which in turn are defined in terms of still more functions, until at the bottom level the functions are language primitives. These functions are much like ordinary mathematical functions, and in this paper will be defined by ordinary equations.'' \citep[p. 1]{hughesWhyFunctionalProgramming1989}}.
Cela distingue la programmation fonctionnelle d'autres paradigmes, notamment en programmation impérative, où l'état est beaucoup plus utilisé, notamment pour le contrôle du flot d'exécution.

Ces propriétés rapprochent beaucoup la programmation fonctionnelle des mathématiques \footnote{%
    ``Functional programming is so called because a program consists entirely of functions. The main program itself is written as a function which recieves the program's input as its argument and delivers the program's output as its result. Typically the main function is defined in terms of other functions, which in turn are defined in terms of still more functions, until at the bottom level the functions are language primitives. These functions are much like ordinary mathematical functions, and in this paper will be defined by ordinary equations.'' \citep[p. 1]{hughesWhyFunctionalProgramming1989}}.

Voici une définition de la programmation fonctionnelle : ``Functional programming is about writing pure functions, about removing hidden inputs and outputs as far as we can, so that as much of our code as possible just describes a relationship between inputs and outputs.'' \citep{jenkisWhatFunctionalProgramming2015}.
Cette définition met l'accent sur le fait que la programmation fonctionnelle cherche à décrire, le plus possible, les programmes comme des relations entre une entrée et une sortie, définie par des fonctions.
L'auteur appelle ``entrées et sorties cachées'' (``hidden inputs and outputs'') toute référence qui ne passe pas par les paramètres ou le retour des fonctions, ce qui inclut notamment les effets de bord.

Les fonctions d'ordre supérieur sont centrales en programmation fonctionnelle.
Les fonctions d'ordre supérieur sont des fonctions qui prennent en argument, ou renvoient d'autres fonctions.
Ces fonctions d'ordre supérieur (aussi appelées ``fonctionnelles'') permettent donc de modifier ou combiner entre elles d'autres fonctions \footnote{%
    ``En mathématiques et en informatique, les fonctions d'ordre supérieur sont des fonctions qui ont au moins une des propriétés suivantes~: elles prennent une ou plusieurs fonctions en entrée~; elles renvoient une fonction.'' \citep{FonctionOrdreSuperieur2023}}.

Pour que les fonctions d'ordre supérieur existent, il est nécessaire que les fonctions soient des citoyens de première classe, c'est-à-dire des entités traitées comme des valeurs du langage.
Il est également nécessaire que le langage implémente les fermetures (voir section \ref{fermeture}), le cas échéant il serait impossible d'implémenter certaines fonctions d'ordre supérieur.
La programmation fonctionnelle privée des fermetures (et donc des fonctions d'ordre supérieur) se nomme la programmation fonctionnelle de premier ordre (voir figure \ref{fig:taxonomie-des-paradigmes}).

\paragraph{Programmation fonctionnelle pure}\label{programmation-fonctionnelle-pure}

La programmation fonctionnelle pure (parfois appelée simplement programmation fonctionnelle) est un paradigme dans lequel toutes les fonctions sont pures, c'est-à-dire qu'il n'existe pas d'état \footnote{%
    La logique combinatoire est un système de calcul dans lequel toutes les fonctions sont pures (il n'y a que des fonctions d'ordre supérieur dans ce système) \citep{LogiqueCombinatoire2023}}.

\subsubsection{Programmation structurée}\label{programmation-structuree}

La programmation structurée peut être définie, du point de vue théorique, comme un paradigme dans lequel le contrôle du flot se fait en utilisant des instructions de contrôle du flot d'exécution particulières (des structures de sélection et d'itération), et qui sont sous la forme de blocks \footnote{%
    ``Structured programming is a programming paradigm aimed at improving the clarity, quality, and development time of a computer program by making extensive use of the structured control flow constructs of selection (if/then/else) and repetition (while and for), block structures, and subroutines.'' \citep{StructuredProgrammingWikipedia}}.
Les structures de sélection sont des structures particulières qui permettent de choisir quel groupe d'instruction sera exécuté en fonction de l'état du programme \footnote{%
    ``\,`Selection'; one or a number of statements is executed depending on the state of the program. This is usually expressed with keywords such as if..then..else..endif. The conditional statement should have at least one true condition and each condition should have one exit point at max.'' \citep{StructuredProgrammingWikipedia}}.
Les structures d'itération sont des instructions qui exécutent répétitivement un block jusqu'à ce que le programme atteigne un certain état, ou bien que certaines opérations aient été appliquées sur chaque élément d'une collection \footnote{%
    ``\,`Iteration'; a statement or block is executed until the program reaches a certain state, or operations have been applied to every element of a collection. This is usually expressed with keywords such as while, repeat, for or do..until. Often it is recommended that each loop should only have one entry point (and in the original structural programming, also only one exit point, and a few languages enforce this).'' \citep{StructuredProgrammingWikipedia}}.

Il est important de voir que le but de la programmation structurée est de produire des programmes plus clairs et de meilleure qualité, de par la modularité induite par la structuration du programme \footnote{%
    ``The most important difference between structured and unstructured programs is that the former are designed in a modular way. Modular design brings great productivity improvements. First of all, small modules can be coded quickly and easily. Secondly, general purpose modules can be reused, leading to faster development of subsequent programs. Thirdly, the modules of a program can be tested independently, helping to reduce the time spent debugging.'' \citep[p. 2]{hughesWhyFunctionalProgramming1989}}.

\subsubsection{Programmation concurrente}\label{programmation-concurrente}

La programmation concurrente permet de gérer des programmes avec plusieurs ensembles d'instructions dont l'exécution doit être indépendante.
En programmation impérative stricte, l'ordre des instructions définit leur ordre d'exécution, ce qui les rend interdépendantes.
On dit alors que ces instructions sont séquentielles.
La programmation concurrente introduit le concept de concurrence, lorsque deux parties d'un programme n'ont pas de dépendance l'une avec l'autre \footnote{%
    ``For example, consider a program that consists of instructions executing one after the other. The instructions are not independent since they are ordered in time. To implement independence we need a new programming concept called concurrency. When two parts do not interact at all, we say they are concurrent.3 (When the order of execution of two parts is given, we say they are sequential.)'' \citep[p. 25]{royProgrammingParadigmsDummies}}.

``Le monde réel est concurrent : il consiste en des activités qui
évoluent indépendamment'' (``The real world is concurrent: it consists
of activities that evolve independently.''
\citep[p. 25]{royProgrammingParadigmsDummies}).

\subsubsection{Les langages multiparadigmes}\label{les-langages-multiparadigmes}

Il existe beaucoup de langages qui implémentent plusieurs paradigmes de programmation \citep{ComparisonMultiparadigmProgramming2024}.
Ces langages sont appelés langages multiparadigmes.

Par exemple, le langage Prolog implémente à la fois la programmation logique et la programmation impérative \footnote{%
    ``Prolog: The ﬁrst paradigm is a logic programming engine based on uniﬁcation and depth-ﬁrst search.The second paradigm is imperative: the assert and retract operations which allow a program to add and remove program clauses.'' \citep[p. 18]{royProgrammingParadigmsDummies}}.
Un autre exemple peut être trouvé lorsque l'on ajoute une librairie de résolution de contraintes à un langage de programmation.
Le paradigme de programmation par contraintes est alors ajouté au langage hôte, qui peut, par exemple, être un langage impératif et orienté objet \footnote{%
    ``Solving libraries (e.g., Gecode): The ﬁrst paradigm is a solver library based on advanced search algorithms, such as Gecode. The second paradigm is added by the host language, e.g., C++ and Java support object-oriented programming.'' \citep[p. 18]{royProgrammingParadigmsDummies}}.

Nous verrons l'avantage des langages de programmation multiparadigmes dans la section \ref{avantages-des-langages-multiparadigmes}.

\subsection{Intérêt intrinsèque des paradigmes}\label{interet-intrinseque-des-paradigmes}

On comprend par leur simple définition un premier intérêt des paradigmes de programmation : ils permettent d'exprimer des concepts différents, de donner des approches différentes sur la programmation.

\subsection{Définition de la puissance d'expression}\label{definition-de-la-puissance-dexpression}

La puissance d'expression (ou expressivité) d'un langage est la quantité et la diversité d'idées qu'il peut représenter ou communiquer \footnote{%
    ``In computer science, the expressive power (also called expressiveness or expressivity) of a language is the breadth of ideas that can be represented and communicated in that language.'' \citep{ExpressivePowerComputer2023}}
\footnote{%
    ``The more expressive a language is, the greater the variety and quantity of ideas it can be used to represent'' \citep{ExpressivePowerComputer2023}}.

On distingue deux sens du terme.
D'un côté, l'expressivité théorique, qui se concentre sur la possibilité théorique d'exprimer une idée dans un langage, indépendamment de la difficulté pour exprimer cette idée.
De l'autre côté, l'expressivité pratique, qui se concentre sur la concision et la facilité d'expression de ces idées \footnote{%
    ``The term expressive power may be used with a range of meaning. It may mean a measure of the ideas expressible in that language: regardless of ease (theoretical expressivity); concisely and readily (practical expressivity)'' \citep{ExpressivePowerComputer2023}}.

\subsubsection{Expressivité au sens formel}\label{expressivite-au-sens-formel}

La puissance d'expression au sens formel (ou expressivité formelle) est mesurée en regardant l'ensemble des idées qu'un langage peut exprimer.
Ce concept d'expressivité formelle est surtout utile en mathématiques ou en informatique théorique (notamment dans la théorie des langages formels) \footnote{%
    ``The first sense dominates in areas of mathematics and logic that deal with the formal description of languages and their meaning, such as formal language theory, mathematical logic and process algebra'' \citep{ExpressivePowerComputer2023}}.

Pour un langage de programmation, il est presque indispensable d'être aussi expressif qu'une machine de Turing (Turing-complets) (ou un autre modèle de calcul équivalent), car cela est la condition nécessaire et suffisante pour qu'il puisse exprimer les problèmes calculables (par définition).

Il existe tout de même des langages qui n'ont pas la même expressivité formelle.
On peut citer les langages de description de données, dont le but est de représenter des données organisées (par exemple, les langages XML, JSON, YAML\ldots), ou bien les langages de description d'ontologies, qui servent à représenter de façon informatique certains types d'ontologie.
Notamment, OWL2 EL et OWL2 RL sont deux langages de description d'ontologies qui n'ont pas la même expressivité formelle : OWL2 EL n'implémente pas certains concepts qui peuvent pourtant être exprimés dans OWL2 RL \footnote{%
    ``For example, the Web Ontology Language expression language profile (OWL2 EL) lacks ideas (such as negation) that can be expressed in OWL2 RL (rule language). OWL2 EL may therefore be said to have less expressive power than OWL2 RL. These restrictions allow for more efficient (polynomial time) reasoning in OWL2 EL than in OWL2 RL. So OWL2 EL trades some expressive power for more efficient reasoning (processing of the knowledge representation language).'' \citep{ExpressivePowerComputer2023}}.

Cependant, ces langages ne sont généralement pas qualifiés de ``langages de programmation'', puisqu'ils sont incapables d'exprimer un programme (c'est-à-dire des instructions exécutables), mais expriment seulement des données.

Il existe également des langages de programmation qui ne sont pas Turing-complets (par exemple, le langage G-code, dans certaines implémentations \footnote{%
    ``G-code began as a limited language that lacked constructs such as loops, conditional operators, and programmer-declared variables with natural-word-including names (or the expressions in which to use them). It was unable to encode logic but was just a way to''connect the dots'' where the programmer figured out many of the dots' locations longhand.'' \citep{Gcode2023}}),
mais ceux-ci ne sont pas utilisés pour résoudre des problèmes de programmation : ils sont utilisés à des fins théoriques, ou pour des applications particulières (le langage G-code est utilisé pour commander des machines à commande numérique \footnote{%
    ``G-code (also RS-274) is the most widely used computer numerical control (CNC) and 3D printing programming language.It is used mainly in computer-aided manufacturing to control automated machine tools, as well as for 3D-printer slicer applications. The G stands for geometry. G-code has many variants.'' \citep{Gcode2023}}).

\subsubsection{Expressivité au sens pratique}\label{expressivite-au-sens-pratique}

L'expressivité pratique est la diversité et la quantité d'idées qu'un langage de programmation peut exprimer facilement.
Dans ce sens de l'expressivité, la facilité d'exprimer une idée est primordiale.
Comme il est dit dans la section \ref{expressivite-au-sens-formel}, la plupart des langages de programmation ont la même expressivité théorique.
Cependant, il n'est pas suffisant pour l'expressivité pratique qu'une idée soit théoriquement exprimable : il est nécessaire qu'il soit aisé de le faire, que le langage implémente des fonctionnalités directement.
Il est, par exemple, possible de faire de la programmation structurée (voir la section \ref{programmation-structuree}) ou de la programmation fonctionnelle (voir la section \ref{programmation-fonctionnelle}) dans un langage qui ne supporte pas nativement ces paradigmes (par exemple le langage assembleur).
En général, pour tout concept \(x\) implémenté par un langage de programmation \(A\), il sera toujours possible de créer un équivalent de \(x\) dans un langage \(B\) qui ne l'implémente pas nativement (à condition que \(B\) soit Turing Complet).
Cependant, l'expressivité au sens pratique ne prend pas en compte cette possibilité, et ne considère que les idées directement (ou presque directement) implémentées et exprimables dans un langage.

L'expressivité pratique d'un langage est donc très liée aux paradigmes qu'il implémente.
On pourrait même définir l'expressivité au sens pratique d'un langage comme la quantité de paradigmes qu'il implémente.

\subsubsection{Compromis entre expressivité et analysabilité}\label{compromis-entre-expressivite-et-analysabilite}

Plus un langage est expressif, plus il est complexe de l'analyser mathématiquement.
Plus un langage est puissant expressivement au sens formel, plus il devient difficile (voire impossible) de démontrer certains théorèmes sur ce formalisme \footnote{%
    ``The design of languages and formalisms involves a trade-off between expressive power and analyzability. The more a formalism can express, the harder it becomes to understand what instances of the formalism say. Decision problems become harder to answer or completely undecidable.'' \citep{ExpressivePowerComputer2023}}.
Cela est vrai également pour l'expressivité au sens pratique : ``le fait d'éviter certaines techniques peut permettre de rendre plus aisée la démonstration de théorèmes sur la correction d'un programme -- ou simplement la compréhension de son fonctionnement~-- sans limiter la généralité du langage de programmation.'' \citep{ParadigmeProgrammation2023}.

\paragraph{Exemple de compromis : automates et grammaires.}\label{exemple-de-compromis-automates-et-grammaires}

Les automates à pile et les machines de Turing forment un exemple de compromis entre analysabilité et expressivité formelle. Les automates à pile reconnaissent les langages non contextuels \footnote{%
    ``{[}\ldots{]} les langages algébriques, appelés aussi langages hors contexte, langages acontextuels, ou langages non contextuels. Ils sont reconnus par un automate à pile.'' \citep{HierarchieChomsky2023}}.
Les machines de Turing reconnaissent les langages contextuels \footnote{%
    ``Les langages {[}\ldots{]} contextuels ou sensibles au contexte, sont exactement ceux reconnus par une machine de Turing non déterministe à mémoire linéairement bornée, appelés couramment automates linéairement bornés.'' \citep{HierarchieChomsky2023}}.
Or, les langages non contextuels sont strictement inclus dans les langages récursivement énumérables \footnote{%
    ``La classe des langages contextuels (type 1) est incluse strictement dans la classe des langages récursivement énumérables (type 0). L'inclusion de la classe des langages algébriques (type 2) dans la classe des langages contextuels (type 1) doit être précisée car un langage contextuel ne contient jamais le mot vide . L'énoncé exact est~: Un langage algébrique ne contenant pas le mot vide est un langage contextuel ou, de manière équivalente~: Un langage algébrique est un langage contextuel éventuellement augmenté du mot vide.'' \citep{HierarchieChomsky2023}}.
On peut donc conclure que les machines de Turing ont un pouvoir d'expression formel supérieur à celui des automates à pile.

Cependant, si on pose le problème de l'appartenance d'un mot à un langage donné : ce problème est décidable pour tous les langages non contextuels \footnote{%
    ``Le problème de l'appartenance d'un mot à un langage algébrique est décidable~: il existe un algorithme qui, étant donnés la description d'une grammaire non contextuelle et un mot, répond en temps fini à la question de l'appartenance de ce mot au langage défini par cette grammaire (plus précisément, on peut le tester en un temps \(O(n^{3})\) pour un mot de longueur n, grâce à l'algorithme CYK).'' \citep{AutomatePile2021}}.
Pourtant, ce problème est indécidable pour les machines de Turing \footnote{%
    ``Le problème de l'appartenance d'un mot à un langage de cette classe {[}la classe des langages récursivement énumérables{]} est indécidable.'' \citep{HierarchieChomsky2023}}.

On voit donc que les machines de Turing sont un formalisme plus expressif (certains concepts exprimés par des machines de Turing ne sont pas exprimables par des automates à pile) mais moins analysable (certains problèmes sont décidables sur les automates à pile, mais pas sur les machines de Turing).

\paragraph{Exemple de compromis : le non-déterminisme}\label{exemple-de-compromis-le-non-determinisme}

Un langage de programmation est dit non déterministe lorsque son exécution ne dépend pas uniquement des spécifications du programme, c'est-à-dire que les spécifications laissent un choix lors de l'exécution du programme \footnote{%
    ``nondeterminism is when the execution of a program is not completely determined by its speciﬁcation, i.e., at some point during the execution the speciﬁcation allows the program to choose what to do next. During the execution, this choice is made by a part of the run-time system called the scheduler'' \citep[p. 14]{royProgrammingParadigmsDummies}}.
Le non-déterminisme est donc problématique, puisqu'il peut amener à créer des programmes dont le résultat est inattendu ou incertain : il rend les programmes moins analysables.
Cependant, certains paradigmes qui peuvent exprimer du non-déterminisme restent utiles pour modéliser certains problèmes (par exemple, la programmation concurrente(voir la section \ref{programmation-concurrente})) \footnote{%
    ``But paradigms that have the power to express observable nondeterminism can be used to model real-world situations and to program independent activities.'' \citep[p. 14]{royProgrammingParadigmsDummies}}.

C'est pourquoi certaines fonctionnalités, certains paradigmes, comme le non-\linebreak{}déterminisme et les paradigmes qui l'impliquent, devraient être utilisés seulement si cela est nécessaire \footnote{%
    ``We conclude that observable nondeterminism should be supported only if its expressive power is needed.'' \citep[p. 14]{royProgrammingParadigmsDummies}}.

\subsubsection{Implications sur la diversité des paradigmes}\label{implications-sur-la-diversite-des-paradigmes}

Nous avons montré qu'il est nécessaire de faire des choix dans les concepts qu'implémente un paradigme, afin de faire un compromis entre analysabilité et expressivité.
Cela donne une justification à l'existence de différents paradigmes : l'implémentation ou non de certains concepts a de réelles influences, non seulement sur l'expressivité des langages, mais également sur les propriétés de ceux-ci (comme le déterminisme).
Il est donc important de pouvoir choisir quels concepts utiliser selon les situations, afin de trouver le meilleur compromis entre expressivité et analysabilité.

%% file: tex_files/2-apprentissage.tex
\section{Paradigmes dans
l'apprentissage}\label{paradigmes-dans-lapprentissage}

\subsection{Importance des paradigmes dans
l'apprentissage}\label{importance-des-paradigmes-dans-lapprentissage}

Les paradigmes de programmation jouent un rôle dans l'apprentissage d'un langage.
Théoriquement, les mêmes idées pourraient être représentées indépendamment du paradigme, mais le paradigme fournit un modèle de pensée, une approche, un cadre.
Ces modèles de pensée sont utiles pour apprendre, car le fait d'avoir en tête des modèles, des idées générales sur la façon d'envisager un programme, permet d'apprendre plus aisément \footnote{%
'' - To help people learn is to help them build, in their heads, various kinds of computational models. - This can best be done by a teacher who has, in his head, a reasonable model of what is in the pupil's head.'' \citep[p. 5]{minskyFormContentComputer1970}}.
Les paradigmes sont encore plus importants du point de vue de l'enseignant, qui doit identifier clairement les paradigmes qu'il enseigne pour pouvoir transmettre efficacement les concepts de la programmation \footnote{%
``To the teacher of programming, even more, I say: identify the paradigms you use, as fully as you can, then teach them explicitly. They will serve your students when Fortran has replaced Latin and Sanskrit as the archetypal dead language.'' \citep[p. 9]{floydParadigmsProgramming1979a}}.
Le fait d'apprendre des paradigmes, des modèles mentaux, permet également de se détacher du langage de programmation particulier, et d'acquérir des connaissances générales, qui restent valides même lorsque l'on change de langage. Maîtriser un paradigme est utile pour tous les langages qui implémentent ce paradigme \footnote{%
``To the teacher of programming, even more, I say: identify the paradigms you use, as fully as you can, then teach them explicitly. They will serve your students when Fortran has replaced Latin and Sanskrit as the archetypal dead language.'' \citep[p. 9]{floydParadigmsProgramming1979a}}.

\subsection{Avantages de la diversité}\label{avantages-de-la-diversite}

On peut se demander s'il est intéressant d'apprendre plusieurs paradigmes au travers d'un ou plusieurs langages de programmation, si cette diversité des paradigmes est utile pour l'apprentissage.

Au fil du temps, à force de lire et travailler avec du code, nos compétences en programmation augmentent \footnote{% 
``reading and working with more code, and more types of code, will increase proficiency at programming.'' \citep[p. 81]{brown10ThingsSoftware2023}}
et l'on mémorise beaucoup de cas qui nous aideront à résoudre des problèmes plus intuitivement et à construire une bibliothèque mentale de modèles\footnote{%
    ``Experts build up a mental library of patterns'' \citep[p. 81]{brown10ThingsSoftware2023}}
que n'aura pas un développeur débutant.
Les sciences cognitives montrent que la différence principale entre un expert et un débutant n'est pas sa capacité à raisonner, mais plutôt à reconnaître des motifs qu'il a déjà vus \footnote{%
    ``One key difference between beginners and experts is that experts have seen it all before. Research into chess experts has shown that their primary advantage is their ability to remember and recognize the state of the board.'' \citep[p. 81]{brown10ThingsSoftware2023}}.
Cela montre l'importance de se construire une telle ``bibliothèque mentale de modèles''.

\subsection{Conclusion sur les paradigmes pour l'apprentissage}\label{conclusion-sur-les-paradigmes-pour-lapprentissage}

Dans le contexte de l'apprentissage de la programmation, les paradigmes présentent plusieurs intérêts :

\begin{itemize}
    \item fournir des modèles mentaux à ceux qui apprennent, afin de mieux envisager la programmation
    \item fournir des des connaissances utiles indépendamment du langage.
    \item fournir des abstractions, des motifs reconnaissables, qui finissent par former une ``bibliothèque mentale de modèles''
\end{itemize}

%% file: tex_files/3-resolution-problemes.tex
\section{Paradigmes pour la résolution de problèmes}\label{paradigmes-pour-la-resolution-de-problemes}

\subsection{Les paradigmes fournissent un cadre pour la pensée}\label{les-paradigmes-fournissent-un-cadre-pour-la-pensee}

La définition même de paradigme explique déjà en partie l'existence des différents paradigmes : un paradigme donne une façon d'envisager la programmation.
Comme il existe de nombreux types de problèmes, de nombreuses situations à modéliser, il semble normal que de nombreux paradigmes existent pour donner les concepts et outils nécessaires afin d'implémenter efficacement des solutions à ces problèmes.

L'existence de nombreux paradigmes de programmation peut donc être justifiée par diversité des problèmes rencontrés : chaque paradigme permet de répondre à une classe de problèmes précis.
Par exemple, la programmation concurrente (voir la section \ref{programmation-concurrente}) permet de modéliser des situations dans lesquelles deux événements indépendants évoluent en même temps et indépendamment.

L'efficacité peut ici avoir deux sens : l'efficacité pour l'humain, c'est-à-dire l'aisance avec laquelle le développeur pourra implémenter une solution à son problème\,; et l'efficacité lors de l'exécution (efficacité pour la machine), qui dépend du temps et de l'espace mémoire nécessaires à l'exécution du programme.
La nécessité de faire des compromis entre expressivité et analysabilité décris plus haut (voir la section \ref{compromis-entre-expressivite-et-analysabilite}) peut avoir pour conséquence la nécessité de faire des compromis entre efficacité pour l'humain et efficacité pour la machine.
En effet, une plus grande efficacité pour l'humain peut être atteinte par une plus grande expressivité (théorique, mais surtout pratique)\,; or, nous avons vu que cela pouvait mener à une moins grande analysabilité du langage, ce qui implique notamment que moins d'optimisation seront possibles lors de l'exécution d'un programme.

\subsection{Avantages des langages multiparadigmes}\label{avantages-des-langages-multiparadigmes}

Lors de la résolution de problèmes, il peut être utile d'avoir de nombreux modèles à disposition, afin de pouvoir choisir celui qui correspond le mieux au problème actuel.
Si ces modèles sont directement implémentés dans notre langage de programmation (s'il supporte les bons paradigmes), la résolution de notre problème sera beaucoup plus aisée.
``A language should ideally support many concepts in a well-factored way, so that the programmer can choose the right concepts whenever they are needed without being encumbered by the others.'' : Un langage devrait, dans l'idéal, intégrer de manière cohérente un grand nombre de paradigmes, pour permettre au développeur de choisir quels concepts il souhaite utiliser, sans être encombré par les autres \citep[p. 10]{royProgrammingParadigmsDummies}.

Certains langages ne nécessitent pas un grand pouvoir d'expression, car ils répondent à un besoin spécifique (voir la section \ref{expressivite-au-sens-formel}).
Cependant, la plupart des langages ont pour but de résoudre une grande diversité de problèmes et il est donc nécessaire qu'ils permettent de décrire et manipuler aisément un grand nombre de concepts.

\subsection{Créer un paradigme pour chaque type de problème}\label{creer-un-paradigme-pour-chaque-type-de-probleme}

Le principe de l'extension créative (de l'anglais \emph{creative extension principle}) est une méthode qui permet de créer de nouveaux paradigmes.
Elle permet de trouver et d'organiser des concepts utiles à la programmation, pour réellement former un paradigme \footnote{%
    ``Concepts are not combined arbitrarily to form paradigms. They can be organized according to the creative extension principle.'' \citep[p. 16]{royProgrammingParadigmsDummies}}.

L'extension créative part de la constatation qu'un problème nécessite des modifications envahissantes (des modifications dans l'ensemble des contextes du programme) pour être résolu.
Il est alors nécessaire de comprendre cette difficulté (pourquoi cette modification devient-elle envahissante ?), et de trouver un concept plus général, plus fondamental, qui résout cette difficulté, c'est-à-dire qui permet d'éliminer ces modifications envahissantes pour retrouver un programme simple \footnote{%
    ``If the need for pervasive modifications manifests itself, we can take this as a sign that there is a new concept waiting to be discovered. By adding this concept to the language we no longer need these pervasive modifications and we recover the simplicity of the program'' \citep[p. 17]{royProgrammingParadigmsDummies}}.

Par exemple, si l'on cherche à modéliser plusieurs activités indépendantes dans un langage purement impératif, il faut implémenter soi-même plusieurs piles d'exécution, ainsi qu'un ordonnanceur. Les modifications impliquées par ce problème sont envahissantes.
Cette complexité peut être évitée si notre langage implémente un concept (et donc un paradigme) : la concurrence (voir la section \ref{programmation-concurrente}) \footnote{%
    ``If we need to model several independent activities, then we will have to implement several execution stacks, a scheduler, and a mechanism for preempting execution from one activity to another. All this complexity is unnecessary if we add one concept to the language: concurrency.'' \citep[p. 17]{royProgrammingParadigmsDummies}}.
La même logique peut s'appliquer à la gestion d'erreurs dans un programme.
Si l'on veut être capable de gérer les erreurs dans un programme, il faut ajouter des conditions dans le corps de chaque fonction, pour que chacune retourne le code d'erreur jusqu'à l'appel initial.
Les modifications impliquées par ce problème sont envahissantes.
Cette complexité peut être évitée si notre langage implémente un concept (et donc un paradigme) : les exceptions \footnote{%
    ``If we need to model error detection and correction, in which any function can detect an error at any time and transfer control to an error correction routine, then we need to add error codes to all function outputs and conditionals to test all function calls for returned error codes.All this complexity is unnecessary if we add one concept to the language: exceptions.'' \citep[p. 17]{royProgrammingParadigmsDummies}}.

Floyd, dans son papier ``The Paradigms of Programming'', explique une méthode similaire qui lui permet de créer de nouveaux paradigmes : lorsqu'il résout un problème complexe, il essaie d'extraire l'essence de sa solution, de la simplifier pour obtenir une solution aussi directe que possible.
Il cherche ensuite une règle générale qui lui permettrait de résoudre des problèmes semblables.
Cette règle, s'il la trouve, peut être le concept fondateur d'un nouveau paradigme \footnote{%
    ``After solving a challenging problem, ! solve it again from scratch, retracing only the insight of the earlier solution. I repeat this until the solution is as clear and direct as I can hope for. Then I look for a general rule for attacking similar problems, that would have led me to approach the given problem in the most efficient way the first time. Often, such a rule is of permanent value.'' \citep[p. 3]{floydParadigmsProgramming1979a}}.

On peut donc voir chaque paradigme comme la réponse à un problème particulier, à une situation qui serait complexe à modéliser sans ce paradigme.

Il serait même pertinent, de ce point de vue, d'encourager la création de nouveaux paradigmes dès que l'on trouve des problèmes nouveaux qui sont complexes à résoudre avec les paradigmes existants.

\subsection{Les paradigmes comme outil pour la pensée}\label{les-paradigmes-comme-outil-pour-la-pensee}

Connaître un langage de programmation ne permet pas de savoir immédiatement comment résoudre tous les problèmes que l'on risque de rencontrer.
Par exemple, la syntaxe des langages dérivés de LISP est très simple, et peut être apprise en peu de temps.
Cependant, connaître la syntaxe complète et le fonctionnement d'un langage ne permettra pas de résoudre tous les problèmes : il est également nécessaire d'être capable de ``faire le lien'' entre un problème et son expression dans un langage de programmation.
C'est ce lien que les paradigmes de programmation permettent de faire.
Plus précisément, les paradigmes permettent de faire le lien entre un problème et une solution théorique, un modèle conceptuel qui permet ensuite d'implémenter une solution.

On peut notamment opposer les paradigmes et les méthodes.
Une méthode permet de convertir en programme des problèmes déjà reconnus dans le cadre d'un paradigme, d'un écosystème.
La méthode ne s'occupe pas de fournir une solution à un problème, ni un modèle pour ce problème, mais permet de convertir cette solution conceptuelle, ce modèle abstrait, en programme dans un langage particulier.
Au contraire, les paradigmes explicitent plutôt quelle vision le développeur doit avoir, et quels concepts il peut utiliser pour construire son modèle du problème.
Un paradigme donne donc un cadre pour modéliser un problème donné.

\begin{figure}
    \centering
    \includegraphics[scale=0.42]{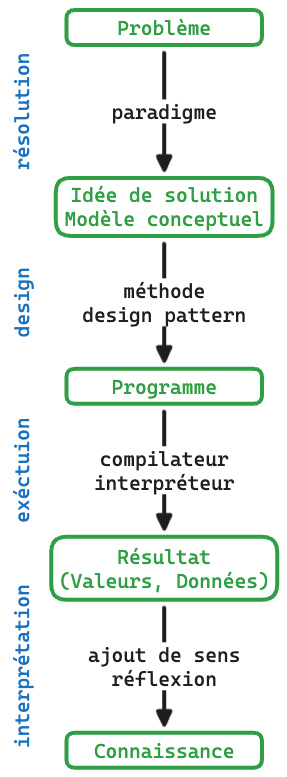}
    \caption{ }
    \label{fig:paradigme-methode-execution-interpretation}
\end{figure}

La figure \ref{fig:paradigme-methode-execution-interpretation} montre les différentes étapes lorsque l'on est
confronté à un problème :

\begin{enumerate}
    \item \textbf{Résoudre} le problème : trouver une idée de solution, et construire conceptuellement un modèle qui permet de résoudre notre problème. Ce sont les paradigmes qui nous permettent de faire cela, en donnant une vision sur ce qu'est un programme, et en fournissant des concepts utiles.
    \item \textbf{Designer} le programme : implémenter concrètement le programme dans un langage de programmation. Ce sont des méthodes, par exemple des design patterns, qui guident le développeur dans cette étape, en lui permettant de convertir des problèmes reconnus en programme dans un langage.
    \item \textbf{Exécuter} le programme : cette étape est réalisée par un compilateur ou un interpréteur.
    \item \textbf{Interpréter} les résultats : pour que les données de sortie du programme deviennent véritablement de la connaissance, il faut leur attacher du sens et un contexte. Il faut pour cela qu'un être humain les interprète. Les statistiques fournissent notamment des outils pour interpréter des séries de données.
\end{enumerate}

Les paradigmes de programmations peuvent donc être vus comme un outil pour la pensée, qui permet --- en fournissant une vision sur la programmation --- de traduire un problème en une pré-solution, en un modèle conceptuel, des modèles mentaux et des concepts utiles.

%% file: tex_files/4-conclusion.tex
\section{Conclusion}\label{conclusion}

En fournissant un modèle pour penser les programmes informatiques, les paradigmes de programmation permettent à la fois d'améliorer l'apprentissage et la résolution de problèmes.
En effet, l'abstraction fournie par les langages est utile à l'enseignement de la programmation (voir la section \ref{paradigmes-dans-lapprentissage}).
Les divers concepts fournis par les différents paradigmes permettent également de modéliser au mieux les différents problèmes.
Pour un programmeur, avoir à sa disposition un grand nombre de paradigmes permet alors de résoudre plus simplement une plus grande variété de problèmes (voir la section \ref{les-paradigmes-fournissent-un-cadre-pour-la-pensee}).
On comprend notamment, dans ce contexte, l'intérêt particulier des langages multiparadigmes, qui laissent au développeur le choix d'utiliser ou non certains concepts selon ses besoins, mais sans avoir à changer de langage de programmation (voir la section \ref{avantages-des-langages-multiparadigmes}).
On peut par ailleurs voir les paradigmes comme résultant de la nécessité de résoudre certaines classes de problèmes.
Un paradigme serait alors à créer pour chaque nouveau type de problème (voir la section \ref{creer-un-paradigme-pour-chaque-type-de-probleme}).
Finalement, on peut comprendre le rôle général des paradigmes dans la résolution de problèmes : permettre la traduction des spécifications et enjeux d'un problème en un modèle conceptuel, en une idée de solution qui est entendable par le développeur (voir la section \ref{les-paradigmes-comme-outil-pour-la-pensee}).

Toutes ces raisons justifient l'existence des nombreux paradigmes de programmation, et encouragent même à en créer de nouveaux.

%% file: tex_files/9-annexes.tex
\section{Annexes}

\begin{figure}[h]
    \centering
    \includegraphics[width=16cm]{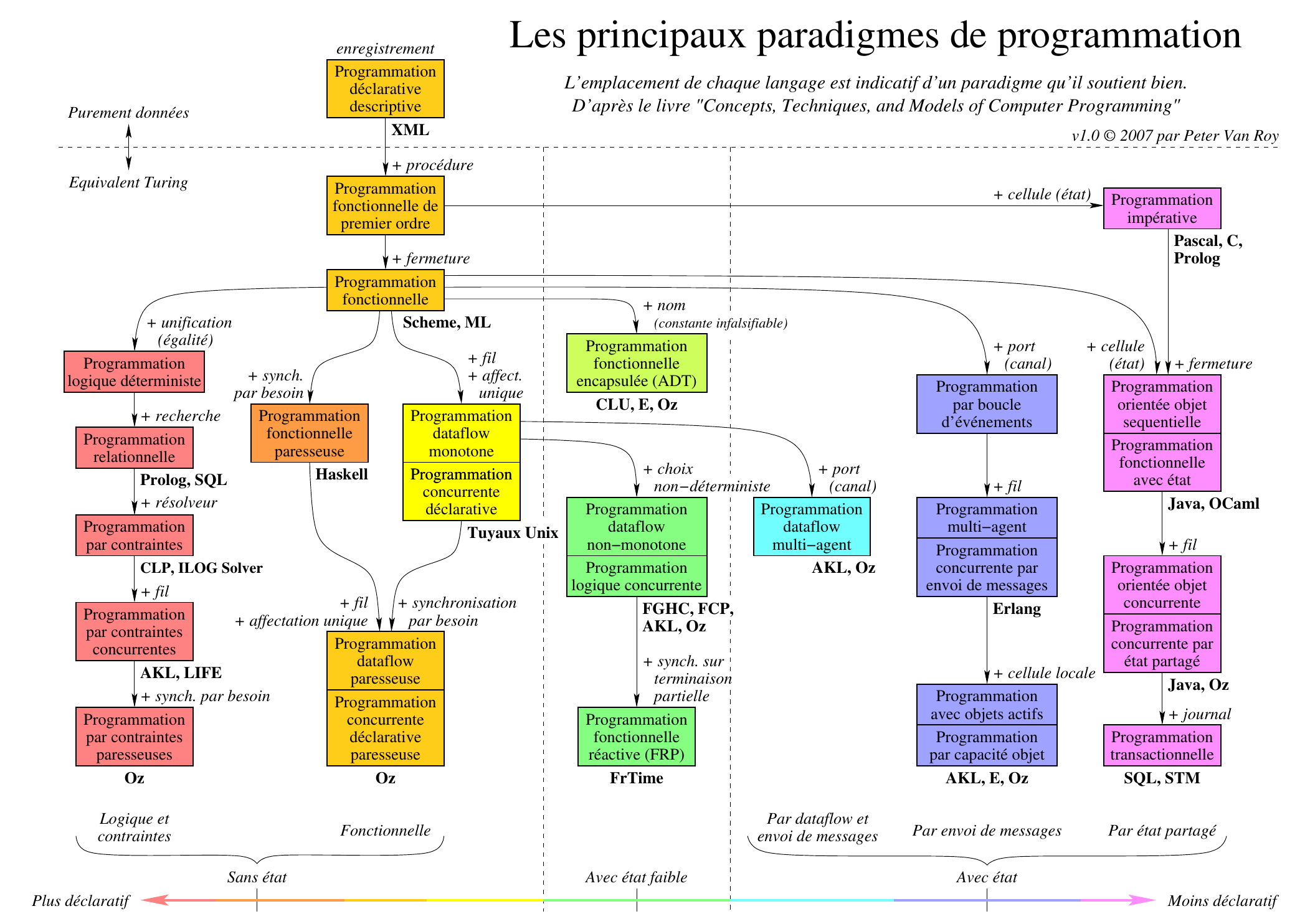}
    \caption{Taxonomie des paradigmes de programmation \cite{TaxonomiePrincipauxParadigmes}}
    \label{fig:taxonomie-des-paradigmes}
\end{figure}